# Community Energy Storage Management for Welfare Optimization Using a Markov Decision Process

Lirong Deng, Xuan Zhang, Tianshu Yang, Hongbin Sun, *Fellow, IEEE*, Shmuel S. Oren, *Life Fellow, IEEE*

*Abstract*—In this paper, we address an optimal management problem of community energy storage in the real-time electricity market under a stochastic renewable environment. In a real-time electricity market, complete market information may not be assessable for a strategic participant, hence we propose a paradigm that uses partial information including the forecast of real-time prices and slopes of the aggregate supply curve to model the price impact of storage use in the price-maker storage management problem. As a price maker, the community energy storage can not only earn profits through energy arbitrage but also smooth price trajectories and further influence social welfare. We formulate the problem as a finite-horizon Markov decision process that aims to maximize the energy arbitrage and social welfare of the prosumer-based community. The advance of the management scheme is that the optimal policy has a threshold structure. The structure has an analytic form that can guide the energy storage to charge/discharge by comparing its current marginal value and the expected future marginal value. Case studies indicate that welfare-maximizing storage earns more benefits than profit-maximizing storage. The proposed threshold-based algorithm can guarantee optimality and largely decrease the computational complexity of standard stochastic dynamic programming.

## I. INTRODUCTION

### A. Background and motivation

Community energy storage is one of the advanced smart grid technologies in recent years, which provides lots of benefits for the electric power system in reliability, quality, economy and control [1]. The community energy storage system could be a single energy storage system or a group of geographically dispersed energy storage systems but coordinated in the form of a virtual power plant. Located close to consumers and distributed energy resources (DERs), community energy storage system acts as an inventory to mitigate the stocahstic DERs and help integrate them into the smart grid [2]. There is an extensive literature investigating the operation, sizing and value of energy storage based on inventory theory [3], [4], especially when facing market uncertainties such as renewable generation, price and demand, e.g., [5]-[7]. These contributions typically assume that energy storage is a price taker that has no influence on market prices.

With the increasing scale of energy storage and market access policies such as Federal Energy Regulatory Commission's Order 841, energy storage is encouraged to participant in the market and participate in price setting, thus acting as a *price maker*. Price-maker energy storage has a complex interaction with the electricity market: on the one hand, the energy storage takes advantage of diffirences in market prices for energy arbitrage; on the other hand, it smoothes the price differences by decreasing on-peak load and increasing off-peak load, resulting in reduction of arbitrage opportunity. To ensure profitablity of the community-owned price-maker energy storage, this paper investigates the management problem from the following perspectives: 1) the interaction between storage operations and uncertain market enviroments; 2) the decision-making structure for energy storage to affect the electricity market; 3) the operational methods beneficial to the community when the community consists of different market participants, such as consumers, generators or prosumers.

### B. Literature review

The difficulty in analyzing the interaction between storage operations and market environments lies in the estimation of uncertain market information, including market prices, aggregate supply curves, market demand, etc. One typical method assumes that supply-demand profiles can be perfectly known for multiple problems, such as planning [8], policy guidance [9], and short-term operation based on games of perfect information [10][11]. An example of the perfect information game is the Stackelberg game, whose mathematical counterpart is the mathematical programming with equilibrium constraints (MPEC). MPEC has high computational cost which may not be applicable in real-time problems. Another typical method to model the impacts of storage operations on market prices is to predict the so-called price quota curve (PQC). The construction of a PQC requires various components of market knowledge, including aggregate supply curves, market prices and demand [12]. In [13], optimal strategies of a price-maker hydro producer were derived with a set of known PQC scenarios. The work in [14] proposed robust optimization for price-maker energy storage to manage the uncertainty risk associated with forecasted PQCs. However, the above methods require extensive market information which is hard to predict in real-time. An alternative is to derive partial but sufficient information to approximately reflect the effects of bidding strategies on market environments, which will be investigated in this paper.

As for the decision-making structures, current practice can be generally categorized into single- and multi-stage decision-making. In the single-stage decision-making, multiple bids are placed simultaneously [11]-[14]. However, more observations of random information will lead to better decisions. Thus, the multi-stage decision-making is appealing in a stochastic environment because each decision is settled with the random information available at that time epoch [5]-[7]. The drawback of the multi-stage decision-making is that it is difficult to solve due to the curse of dimensionality.

Addtionally, to operate in a community-level beneficial way, community energy storage should consider not only energy arbitrage profits but also social welfare gains of the community, reaching a win-win outcome [9], [15], [16]. With the growing scale of energy storage, the welfare benefits become significant, which may stimulate different ownership, such as consumers, producers and prosumers, to focus on their own welfare, thus further influencing storage use. In particular, consumers are likely to increase consumer surplus, so they tend to overuse storage. Conversely, producers may decrease producer surplus, so they tend to underuse storage [9]. However, most studies neglect the potential welfare changes due to storage use. Although references [9], [15], [16] provide an insightful study of welfare impacts, simplified market settings and lack of uncertainty analysis may hinder the application of the models to practical electricity markets.

### C. Contributions

To address the issues mentioned above, this paper focuses on strategic community energy storage and proposes an optimal storage management problem that maximizes both energy arbitrage and community social welfare in face of the uncertainty of renewable energy and real-time price. For clarity, we refer to maximizing energy arbitrage gains as a *profit-maximizing* problem and to maximizing both energy arbitrage gains and social welfare as a *welfare-maximizing* problem. To quantify the effects of community behaviors (i.e., energy consumption, RES generation, storage operation) on price, a novel framework is introduced that only needs to predict partial market information, namely, the real-time market prices, local slopes of the aggregate supply curves around the forecasted market prices, and the slope of the market demand curve, other than knowing the complete market information. We model the uncertainty of renewable generation as a Markovian process and formulate the problem as a discrete-time Markov decision process (MDP) over a finite horizon. The MDP format is a natural choice due to the temporal correlations between storage actions and realizations of random variables in the real-time market setting. Our contributions are summarized as follows:

  1) We propose a multi-stage management scheme of welfare-maximizing community energy storage in a pool-based electricity market under stochastic renewable generations and prices. To the best of the authors' knowledge, this is the first paper that determines strategic behaviors of energy storage in an MDP format considering arbitrage gains and welfare impacts. In addition, the scheme for price-maker participants bidding requires only partial market information, including real-time price forecast, slopes of the aggregate supply curve, and the slope of the market demand curve.

  2) We show a threshold structure of the optimal policy that allows us to determine the optimal charging/discharging rates under dynamic pricing conditions. The threshold compares the current marginal value and the expected future marginal value. With an analytical format, the threshold-based algorithm avoids part of the state discretization as in the standard stochastic dynamic programming (SDP) algorithm.

## II. SYSTEM SETUP

### A. System structure

Suppose that there are some (local) consumers and renewable generators in a community. The community acts as a prosumer participating in the real-time market, which sells/buys power to/from the electric grid at real-time prices. In this study, we focus on the benefits and impacts of energy storage operation, rather than the economic values of excess generation sales or load cost.

Assume that the community wants to invest in energy storage, which has a relatively large scale such that it influences the market price. Fig. 1 shows the interaction among energy storage, the electricity market and social welfare. Note that we differentiate the electricity market environment into ex ante and ex post stages. Ex ante and ex post refer to stages before and after the storage operation, respectively. Such separate descriptions help us identify storage benefits. In this paper, we mainly focus on two benefits, namely, energy arbitrage gains [5] and social welfare gains [9]. Energy arbitrage takes advantage of price differences by buying and storing energy when prices are low and discharging and reselling it when prices are high. On the other hand, when the market price is increased (resp. reduced) through storage charging (resp. discharging), consumer surplus will decrease (resp. increase), and producer surplus will increase (resp. decrease). Notably, although storage belongs to the community, the social welfare of all the other participants in the market will also change due to storage use.

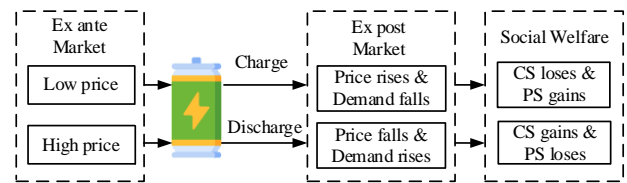

Fig. 1. A general structure of financial impacts among energy storage behaviors, the electricity market and social welfare. CS is short for consumer surplus, while PS is short for producer surplus.

There are three parties in the community: load, renewable energy and energy storage. Assume that renewable energy has zero marginal cost, storage discharging has zero marginal cost, and storage charging has the highest utility. Therefore, there is no difference in community arbitrage profits between considering renewable/storage bidding in the market (front-of-meter) and directly supplying the load (behind-the-meter). However, this is not the case for welfare analysis. Since energy storage itself has no surplus, if its behaviors are behind the meter, community social welfare changes will be miscalculated. Hence, in this paper, we consider renewable, load and energy storage bid in front of meters, that is, they bid separately in the market.

### B. Market setting

  1) **Load**. Denote the community load by $d_t$ and assume that it consists of two parts:

$$d_t = D_t(p_t) = \begin{cases} a_t - b \cdot p_t & p_t \leq p_t^{\max} \\ 0 & p_t > p_t^{\max} \end{cases} \quad (1)$$

where $p_t$ is the market clearing price (MCP) and $a_t$ represents the community maximum load. If MCP is larger than the highest accepex price level $p_t^{\max}$, loads will decrease to zero. $b$ is related to the price elasticity of the community load. $a_t$, $b$>0. Subscript $t$ refers to the time index and will be used throughout the rest of the paper. Similar formulations are applied to the total market load $d_t^{\text{all}}$ with structure $D_t^{\text{all}}$ and parameters $a_t^{\text{all}}$ and $b^{\text{all}}$. We do not need to know all the information of the total market load but only the total price elasticity $b^{\text{all}}$.

2) **Community renewable**. Define $RE_t$ as the total renewable production in the community. Suppose that the renewable generation follows a Markov chain.

3) **Market price**. The market price can be predicted by:
$$p_{0,t} = \tilde{p}_{0,t} + \varepsilon \qquad (2)$$
where $p_{0,t}$ denotes the actual market price value at time $t$. The subscript 0 is defined as the market state without accounting for community participation in the market; $\tilde{p}_{0,t}$ is the predicted price; and $\varepsilon$ is the predicted error which is assumed to follow a Gaussian distribution with zero mean and standard variance. Suppose the aggregate supply (price-quantity) function has the form $p_t = J(g_t)$. The shape of $J(\cdot)$ depends on the market structure and generator offers. It can be (approximately) polynomial, piecewise linear, or even nonconvex. Accurately estimating $J(\cdot)$ is difficult. We propose a paradigm that only needs to know the local slopes of the supply curve near the forecasted price, thus quantitatively estimating the effects of storage charging/discharging behaviors on market prices. An intuitive yet useful idea is to make the first-order Taylor expansion around the predicted value $\tilde{p}_{0,t}$:
$$p_t \approx J(\tilde{g}_{0,t}) + J'(\tilde{g}_{0,t})(g_t - \tilde{g}_{0,t}) = \tilde{e}_t + \tilde{h}_t \cdot g_t \qquad (3)$$
where $\tilde{g}_{0,t} = J^{-1}(\tilde{p}_{0,t})$, $\tilde{e}_t$ and $\tilde{h}_t$ are the intercept and slope of the local linearization of the aggregate supply curve around $\tilde{p}_{0,t}$, respectively. We assume that $J(\cdot)$ is non-decreasing, thus, $\tilde{h}_t \geq 0$.

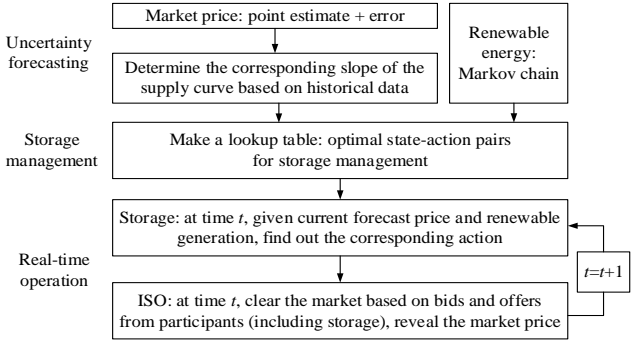

Fig. 2. The control scheme of the energy storage in the real-time market.

Note that in the electricity market, suppliers are mainly traditional generators and RESs. Since the marginal prices of traditional units are higher than those of RESs, the aggregate supply curve can be divided into two parts according to the prices: the RES-aggregation part and the traditional generator-aggregation part. Due to the high uncertainty of RES, the RES-aggregation part changes greatly. For traditional units, on the one hand, the bidding strategies are more stable; on the other hand, because there are many generators in the market, the strategic change made by a single unit has a relatively small effect on the traditional generator-aggregation part. Thus, the sensitivity of this part to the price can be regarded unchanged. That is, the slope of the local linearized curve $\tilde{h}_t$ is approximately constant [17]. In contrast, the intercept of the local linearized curve $\tilde{e}_t$ is stochastic due to the shifts of the RES-aggregation part. Fortunately, we only require knowing $\tilde{h}_t$ in this paper, which addresses the relationship between a quantity change of a supplier and the resulting price change. $\tilde{h}_t$ can be estimated through historical market data.

Based on the proposed system setup, we present a control scheme of energy storage illustrated in Fig. 2. There are three parts in the scheme: at the beginning, the uncertainty forecasting is conducted based on historical data. Then, the storage management is presented to derive a lookup table of optimal state-action pairs. After obtaining the lookup table, the storage owner operates the storage in the real-time market.

III. OPTIMAL STORAGE MANAGEMENT

In this section, we first illustrate the energy storage model and analyze the social welfare gain and then formulate the MDP of the storage management problem. Finally, we provide an SDP method to solve the problem.

*A. Energy storage model*

The energy storage model can be characterized by the following metrics:

1) Energy capacity: the size of storage, denoted by $C$.
2) Round trip efficiency: the ratio of the energy discharged to the demand to the energy charged from the system over each cycle. In this paper, two specific conversion loss efficiencies are used; namely, charging efficiency $\eta_c$ and discharging efficiency $\eta_d$. $\eta_c, \eta_d \in (0,1)$.
3) Capital cost: the investment cost of storage per unit of storage capacity.

Denote $u_t$ as the charging power and $w_t$ as the discharging power. Due to conversion losses, $\eta_c^{-1} u_t$ MW of energy should be purchased from the market to ensure that $u_t$ MW is stored in storage. Correspondingly, $\eta_d w_t$ MW of energy can be received by the grid if the storage discharges $w_t$ MW.

*B. Social welfare gains*

For clarity, we divide the community behaviors into four possible states: 1) initial state: only load bids; 2) RES state: load and RES bid; 3) storage charging state; and 4) storage discharging state. In particular, the price in the initial market state 0 is predicted without the operations of community renewable energy and storage. In what follows, we use subscripts $c$ and $d$ to represent charging and discharging states, respectively. Under each state, the market equilibrium is the intersection of the total demand curve and the aggregate supply curve.

1) $p_{0,t} = (D_t^{\text{all}})^{-1}(d_{0,t}^{\text{all}}) = \tilde{e}_t + \tilde{h}_t \cdot g_{0,t}, \ g_{0,t} = d_{0,t}^{\text{all}}$ (4)

2) $p_t = (D_t^{\text{all}})^{-1}(d_t^{\text{all}}) = \tilde{e}_t + \tilde{h}_t \cdot g_t, \ g_t = d_t^{\text{all}} - RE_t$ (5)

3) $p_{c,t} = (D_t^{\text{all}})^{-1}(d_{c,t}^{\text{all}}) = \tilde{e}_t + \tilde{h}_t \cdot g_{c,t}, \ g_{c,t} = d_{c,t}^{\text{all}} - RE_t + \eta_c^{-1} u_t$ (6)

4) $p_{d,t} = (D_t^{\text{all}})^{-1}(d_{d,t}^{\text{all}}) = \tilde{e}_t + \tilde{h}_t \cdot g_{d,t}, \ g_{d,t} = d_{d,t}^{\text{all}} - RE_t - \eta_d w_t$ (7)

Since we focus our attention on storage management, market prices $p_{c,t}$ and $p_{d,t}$ are defined as the ex post prices, while $p_t$ can be regarded as ex ante information. The expressions of $p_t$, $p_{c,t}$ and $p_{d,t}$ are given as

$$p_t = p_{0,t} - \tilde{h}_t RE_t / (1 + b^{\text{all}} \tilde{h}_t) \qquad (8)$$

$$p_{c,t} = p_t + \tilde{h}_t \eta_c^{-1} u_t / (1 + b^{\text{all}} \tilde{h}_t) \qquad (9)$$

$$p_{d,t} = p_t - \tilde{h}_t \eta_d w_t / (1 + b^{\text{all}} \tilde{h}_t). \qquad (10)$$

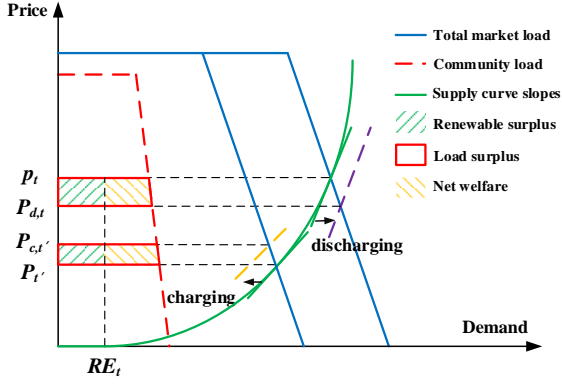

Fig. 3. Social welfare changes due to storage use. Charging behavior of the storage will lift the price, thus decreasing the consumer surplus, but increasing the renewable surplus. Discharging helps lower the price, increase the consumer surplus and decrease the renewable surplus. $p_t$ refers to high ex ante price, while $p_{t'}$ refers to low ex ante price.

According to [9], changes in market prices and total load will result in welfare changes for both consumers and generators in charging and discharging states. Fig. 3 depicts the social welfare changes of the prosumer-oriented community. Note that Fig. 3 only shows when demand is larger than renewable energy. A similar analysis could be conducted when renewable energy plays a dominant role. We are interested in the storage effects on net welfare gains $W_c(u_t, RE_t)$ and $W_d(w_t, RE_t)$ of the community during charging and discharging states:

$$W_c(u_t, RE_t) = RE_t \cdot (p_{c,t} - p_t) - \int_{p_t}^{p_{c,t}} D_t(p)dp$$
$$= -(a_t - bp_t - RE_t)\frac{\tilde{h}_t \eta_c^{-1}}{1+b^{\text{all}}\tilde{h}_t} u_t + \frac{1}{2}b\left(\frac{\tilde{h}_t \eta_c^{-1}}{1+b^{\text{all}}\tilde{h}_t}\right)^2 u_t^2 \quad (11)$$

$$W_d(w_t, RE_t) = -RE_t \cdot (p_t - p_{d,t}) + \int_{p_{d,t}}^{p_t} D_t(p)dp$$
$$= (a_t - bp_t - RE_t)\frac{\tilde{h}_t \eta_d}{1+b^{\text{all}}\tilde{h}_t} w_t + \frac{1}{2}b\left(\frac{\tilde{h}_t \eta_d}{1+b^{\text{all}}\tilde{h}_t}\right)^2 w_t^2. \quad (12)$$

### C. Problem formulation

The energy storage is operated periodically. To avoid plundering opportunity cost from the next period, it is commonly assumed that energy storage should return to its initial storage level at the end of the last stage of the current scheduling period:

$$X_{T+1} = X_1 \quad (13)$$

where $X$ is the storage level and $T$ is the number of stages in a period. The charging action $u_t$ and discharging action $w_t$ are limited by:

$$0 \leq u_t \leq C - X_t, \quad (14)$$
$$0 \leq w_t \leq X_t. \quad (15)$$

The charging and discharging behaviors of energy storage cannot be conducted simultaneously. Hence, there is a complementary constraint $u_t \cdot w_t = 0$, which makes the problem strongly nonconvex. However, if we take conversion losses into consideration, according to [18], the complementary constraint can be eliminated directly without sacrificing the feasibility and optimality of the original problem. The intuition is that charging and discharging storage simultaneously is not economic if considering conversion loss.

The welfare optimization problem in the real-time market is a sequential decision-marking problem, which can be characterized as an MDP.

In our MDP format, system states include 1) endogenous state variable, storage level $X$, and 2) exogenous state variable, renewable generation $RE$. In this paper, we assume that the transition of $RE$ is Markovian, which captures the probability distribution and temporal correlation of the renewable generation. Let $f(X_{t+1}, RE_{t+1} \mid X_t, RE_t, u_t, w_t)$ denote the probability mass function of storage level $X_{t+1}$ and renewable production $RE_{t+1}$, given particular values of the preceding storage level $X_t$, renewable production $RE_t$, and action $u_t$ and $w_t$. Note that $f(X_{t+1}, RE_{t+1} \mid X_t, RE_t, u_t, w_t) = f(RE_{t+1} \mid RE_t)$. The reason is twofold: one is that, given storage level $X_t$ and actions $u_t$ and $w_t$, the next storage level $X_{t+1}$ has been determined through

$$X_{t+1} = X_t + u_t - w_t; \quad (16)$$

the other is that the renewable generation $RE_t$ does not rely on storage level $X_t$ or action $u_t$, $w_t$.

The reward function $r_t$ includes arbitrage profits and net welfare gains under charging and discharging states:

$$r_t(u_t, w_t, RE_t) = g_d(w_t, RE_t) - g_c(u_t, RE_t) \quad (17)$$

where

$$g_d(w_t, RE_t) = \mathbb{E}_{p_t}(\eta_d w_t \cdot p_{d,t} + W_d(w_t, RE_t))$$
$$= \mathbb{E}_{p_t}\left(\eta_d p_t + \tilde{h}_t \eta_d (a_t - RE_t - bp_t)/(1+b^{\text{all}}\tilde{h}_t)\right) w_t$$
$$- \left(\frac{\tilde{h}_t \eta_d^2}{1+b^{\text{all}}\tilde{h}_t} - \frac{1}{2}b\left(\frac{\tilde{h}_t \eta_d}{1+b^{\text{all}}\tilde{h}_t}\right)^2\right) w_t^2$$

$$g_c(u_t, RE_t) = \mathbb{E}_{p_t}(\eta_c^{-1} u_t \cdot p_{c,t} - W_c(u_t, RE_t))$$
$$= \mathbb{E}_{p_t}\left(\eta_c^{-1} p_t + \tilde{h}_t \eta_c^{-1}(a_t - RE_t - bp_t)/(1+b^{\text{all}}\tilde{h}_t)\right) u_t$$
$$+ \left(\frac{\tilde{h}_t \eta_c^{-2}}{1+b^{\text{all}}\tilde{h}_t} - \frac{1}{2}b\left(\frac{\tilde{h}_t \eta_c^{-1}}{1+b^{\text{all}}\tilde{h}_t}\right)^2\right) u_t^2.$$

After analyzing the above actions, states, and the reward function, the storage management problem can be formulated as follows:

$$\max_{u_t, w_t} \mathbb{E}\left(r_N + \sum_{t=1}^{T-1} r_t\right) \quad (18)$$
$$s.t. \quad (13)\text{-}(16).$$

The corresponding Bellman equations are given by

$$V_t(X_t, RE_t) = \max_{u_t, w_t} r_t(u_t, w_t, RE_t)$$
$$+ \mathbb{E}_{RE_{t+1}} V_{t+1}(X_{t+1}, RE_{t+1}) \quad (19)$$

$$V_T(X_T, RE_T) = \begin{cases} r_T(X_1 - X_T, 0, RE_T) & X_1 > X_T \\ r_T(0, X_T - X_1, RE_T) & X_1 \leq X_T \end{cases}$$

where renewable process $RE_t$ is a Markovian process. $V_t(X_t, RE_t)$ is the optimal expected return of optimal action $u_t$ and $w_t$ in the storage level $X_t$ and renewable state $RE_t$ for a finite-horizon problem starting at time $t$ and ending at time $T$. $\mathbb{E}[\cdot]$ denotes expectation. $V_T(X_T, RE_T)$ is the terminal profit. If $X_1 > X_T$, the storage at time $T$ is charged to $X_1$, i.e., $u_T = X_1 - X_T$, $w_T = 0$; otherwise, the storage is discharged to $X_1$, i.e., $w_T = X_T - X_1$, $u_T = 0$.

## D. Stochastic dynamic programming

We solve the MDP model by SDP on a discretized state space. Storage levels and renewable states are discretized into $N_{SOC}$ and $N_{RES}$ intervals, respectively. The pseudocode of the SDP is summarized in Algorithm 1.

**Algorithm 1: Stochastic dynamic programming**
```
1:  X₁ ← const, {REₜ}, {f(REₜ₊₁| REₜ)}
2:  for t=T-1 to 1 do
3:    for Xₜ=1 to N_SOC states do
4:      for REₜ=1 to N_RES states do
5:        for Xₜ₊₁=1 to N_SOC states do
6:          φₜ₊₁ ← 0
7:          for REₜ₊₁=1 to N_RES states do
8:            if t==T-1 then
9:              if X₁>X_T then
10:                V_T(X_T,RE_T)=r_T(X₁-X_T,0,RE_T)
11:             else
12:                V_T(X_T,RE_T)=r_T(0,X_T-X₁,RE_T)
13:             end if
14:           end if
15:           φₜ₊₁ ← φₜ₊₁+f(REₜ₊₁| REₜ)V*ₜ₊₁(Xₜ₊₁,REₜ₊₁)
16:         end for
17:         if Xₜ₊₁<Xₜ then
18:           wₜ ← Xₜ-Xₜ₊₁, uₜ ← 0
19:           evaluate Gₜ(uₜ,wₜ,Xₜ,REₜ) = rₜ + φₜ₊₁
20:         else
21:           uₜ ← Xₜ-Xₜ₊₁, wₜ ← 0
22:           evaluate Gₜ(uₜ,wₜ,Xₜ,REₜ) = rₜ + φₜ₊₁
23:         end if
24:       end for
25:       V*ₜ(Xₜ,REₜ) = max_{uₜ,wₜ} Gₜ(uₜ,wₜ,Xₜ,REₜ),
26:       {u*ₜ(Xₜ,REₜ),w*ₜ(Xₜ,REₜ)} = arg max_{uₜ,wₜ} Gₜ(uₜ,wₜ,Xₜ,REₜ).
27:     end for
28:   end for
29: end for
```

## IV. OPTIMAL THRESHOLD STRUCTURE

Finite horizon models using SDP suffer from the "curse of dimensionality", which makes them computationally intractable to obtain optimal solutions; that is, in order to obtain the optimal solution by backward induction, the states need to be highly discretized, in that the complexity of backward induction grows exponentially as the size of states increases. To tackle this problem, in this section, we show that the reward function is concave in $u$ and $w$, while the cost-to-go function is concave in $X_t$. Then, the optimal solution has a threshold structure, which reduces the discretization in the backward induction process from $V_{t+1}$ to $V_t$.

*Lemma 1:* For every $u_t$, $w_t$ and $t=1,\ldots,T$, $g_d(w_t, RE_t)$ is a concave function in $w_t$, and $g_c(u_t, RE_t)$ is a convex function in $u_t$. $V_t(X_t, RE_t)$ is a concave function in $X_t$.

*Proof:* It is observable that both $g_d(w_t, RE_t)$ and $g_c(u_t, RE_t)$ are quadratic functions; thus, it is easy to analyze the convexity or concavity property. Specifically, we need to deduce the positivity or negativity of the quadratic terms, which is equal to distinguishing the symbol of $\pm(1+b^{all}\tilde{h}_t - 0.5b\tilde{h}_t)$. Note that the slope of the supply curve $\tilde{h}_t$ is nondecreasing; i.e., $\tilde{h}_t \geq 0$, which indicates that higher prices make power more profitable to produce. In addition, we know that $b^{all} > b$ by definition. Hence, $g_d(w_t, RE_t)$ is concave, $g_c(u_t, RE_t)$ is convex, and $r_t(u_t, w_t, RE_t)$ is concave. We use induction to prove the concavity of $V_t(X_t, RE_t)$. 1) The terminal profit $V_T(X_T, RE_T)$ is a concave function by definition. 2) Suppose that $V_{t+1}(X_{t+1}, RE_{t+1})$ is concave for any $t < T-1$. This implies that $\mathbb{E}_{RE_{t+1}} V_{t+1}(X_{t+1}, RE_{t+1})$, as the weighted mean of $V_{t+1}(X_{t+1}, RE_{t+1})$, is also concave in $X_{t+1}$. Then, $V_t(X_t, RE_t)$ is concave in $X_t$. 1) and 2) conclude the proof. Q.E.D.

Denote $h_t(X_t, RE_t) \triangleq \partial V_t/\partial X_t$, which represents the expected marginal value starting from time $t$. For notation simplicity, we use $h_{t+1}(X_{t+1})$ to represent the expected future marginal value of $h_{t+1}(X_{t+1}, RE_{t+1})$ over $RE_{t+1}$, i.e., $\mathbb{E}_{RE_{t+1}} h_{t+1}(X_{t+1}, RE_{t+1})$. In the following Theorem 1, we derive a threshold structure of the optimal policy based on Lemma 1. The threshold is to compare the derivative of $g_d(w_t, RE_t)$ or $g_c(u_t, RE_t)$ with $h_{t+1}(X_{t+1})$ for a given $RE_t$.

*Theorem 1:* The threshold structure of the optimal policy for stage $t$ is

$$\left(w_t^*(X_t, RE_t), u_t^*(X_t, RE_t)\right) =$$
$$\begin{cases} (X_t, 0), & \text{if } \partial g_d(X_t)/\partial w_t \geq h_{t+1}(0) \\ (0, C-X_t), & \text{if } \partial g_c(C-X_t)/\partial u_t \leq h_{t+1}(C) \\ (x, 0), & \text{if } \partial g_d(0)/\partial w_t > h_{t+1}(X_t) \text{ and } \partial g_d(X_t)/\partial w_t < h_{t+1}(0) \\ \quad \text{where } x = \sup\{x: \partial g_d(x)/\partial w_t \geq h_{t+1}(X_{t+1})\} \\ (0, y), & \text{if } \partial g_c(0)/\partial u_t < h_{t+1}(X_t) \text{ and } \partial g_c(C-X_t)/\partial u_t > h_{t+1}(C) \\ \quad \text{where } y = \sup\{y: \partial g_c(y)/\partial u_t \leq h_{t+1}(X_{t+1})\} \\ (0, 0), & \text{if } \partial g_c(u_t)/\partial u_t \geq h_{t+1}(X_{t+1}) \text{ and } \partial g_d(w_t)/\partial w_t \leq h_{t+1}(X_{t+1}). \end{cases}$$
(20)

*Proof sketch:* Lemma 1 shows that Problem (18) is a concave problem with linear constraints. Therefore, KKT conditions are necessary and sufficient conditions for optimality. By deriving the KKT conditions of (19), we have Theorem 1.

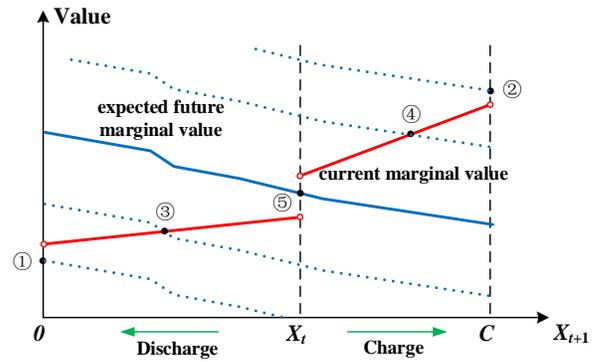

Fig. 4. Expected future marginal value $h_{t+1}$ vs. current marginal value $\partial g_d/\partial w_t$ or $\partial g_c/\partial u_t$. Redline segments are current marginal values. Blue lines and dots represent expected future marginal value curves in five possible conditions. The optimal control $u_t^*, w_t^*$ is depicted by black spots and marked as ①②③④⑤, which are illustrated in (18) in sequence.

Optimal control $u_t^*, w_t^*$ depends on the balance between the expected future marginal value and the current marginal value, as shown in Fig. 4. Charging behavior will increase the current marginal value but decrease the expected future marginal value; discharging behavior is the opposite.

Scenario ① tells us that the current marginal value is no less than the expected future value, even if the energy storage is fully discharged, then the energy storage will prefer the current profits and thus discharge to empty. In contrast, if the current marginal value remains lower than the expected future marginal value even in the fully charged situation, the storage will charge to the full state to obtain enough reserves for future sale, which is Scenario ②. If the future value intersects with the current marginal value; i.e., in Scenarios ③④, the storage will charge/discharge to the storage state indicated by the intersection point. Otherwise, there will be no operation under Scenario ⑤.

We can directly obtain $h_t(X_t, RE_t)$ through KKT conditions of (19),

$$h_t(X_t, RE_t) = \begin{cases} \frac{\partial g_d(X_t)}{\partial w_t}, & \text{if } \frac{\partial g_d(X_t)}{\partial w_t} \geq h_{t+1}(0) \\ \frac{\partial g_c(C-X_t)}{\partial u_t}, & \text{if } \frac{\partial g_c(C-X_t)}{\partial u_t} \leq h_{t+1}(C) \\ h_{t+1}(X_{t+1}), & \text{otherwise.} \end{cases} \quad (21)$$

Note that (20) lists five scenarios, whereas (21) has three scenarios, which summarizes the last three from (20) into one. The pseudocode of the threshold-based algorithm is shown in Algorithm 2.

Comparing Algorithm 2 with Algorithm 1, the SDP algorithm has two discretization processes of the storage level, namely, the current level $X_t$ and the next storage level $X_{t+1}$. The threshold-based algorithm has one discretization process of the current level $X_t$ and removes the discretization process from $V_{t+1}$ to $V_t$ by using the optimal solution structure, which can reduce the computational complexity. Therefore, the proposed algorithm will accelerate the SDP process.

| Algorithm 2: Threshold-based algorithm |
|---|
| 1: $X_1 \leftarrow$ const, $\{RE_t\}$, $\{f(RE_{t+1}\|RE_t)\}$ |
| 2:   **for** $t=T$ **to** 1 **do** |
| 3:     **for** $X_t=1$ **to** $N_{SOC}$ states **do** |
| 4:       **for** $RE_t=1$ **to** $N_{RES}$ states **do** |
| 5:         **if** $t==T$ **then** |
| 6:           **if** $X_1>X_t$ **then** |
| 7:             $h_t(X_t, RE_t) = \partial g_c(X_1-X_t)/\partial u_t$ |
| 8:           **else** |
| 9:             $h_t(X_t, RE_t) = \partial g_d(X_t-X_1)/\partial w_t$ |
| 10:           **end if** |
| 11:         **end if** |
| 12:         Compute $w_t^*(X_t, RE_t), u_t^*(X_t, RE_t)$ using (20) |
| 13:         Compute $h_t(X_t, RE_t)$ using (21) |
| 14:         Take the integral of $h_t(X_t, RE_t)$ over $RE_t$ to get $h_t(X_t)$ |
| 15:       **end for** |
| 16:     **end for** |
| 17:   **end for** |

## V. CASE STUDIES

In this section, we use numerical examples to demonstrate our theoretical results. There are three cases for comparison:

Case 1: regard storage as a price taker.

Case 2: regard storage as a profit-maximizing strategic resource. That is, the objective considers energy arbitrage.

Case 3: regard storage as a welfare-maximizing strategic resource. That is, the objective includes energy arbitrage and social welfare.

Assume that the charging/discharging efficiency of the invested storage is 0.9. The rated capacity of energy storage is 20 MWh. The number of intervals of the energy storage level is 20. The forecasted real-time prices were chosen from [20] during Sep. 1 to Sep. 24, 2020. Supply curve slopes were based on [21]; we scale them in Table I. Price elasticity for total market load was chosen to be $b^{\text{all}}=0.5$. The parameters for community load were set as $a_t^0 = 10$, $b=0.2$. Based on the data set from [22], we took the maximum load of 11 residential consumers as community maximum load $a_t$ and the output of wind turbine 13 multiplied by a scale parameter 4 as the renewable production $RE_t$. To match the real-time price, we interpolated the load and renewable energy into 5-minute granularity. Real-time prices and community net load ($a_t - RE_t$) are depicted in Fig. 5.

TABLE I. TYPICAL SLOPES OF THE SUPPLY CURVE IN REAL-TIME MARKET

| Period | 1 | 2 | 3 | 4 | 5 | 6 |
|---|---|---|---|---|---|---|
| Ex ante price | 0~2 | 2~16 | 16~25 | 25~38 | 38~57 | 57~240 |
| Slope | 0.004 | 0.131 | 0.043 | 0.166 | 0.665 | 6.020 |

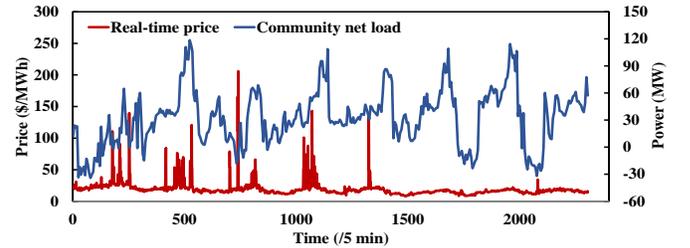

Fig. 5. Real-time prices and community net load profile.

### A. Periodicity and initial storage level settings

Periodicity $T$ refers to the scheduling period of the storage. The initial storage level means the storage level at the beginning of the period, which is also the storage level at the end of the period.

To choose the most suitable values of periodicity and initial storage level for the selected loads, we tested optimal values of different periodicities under different initial storage levels. The number of sampling points in 1/4 day is 72 since the load is collected once every 5 minutes. Finally, the periodicity was set as (1, 2, 3, 4, 5, 6)*72. That is, periodicity is chosen as 0.25, 0.5, 0.75, 1, 1.25, and 1.5 days.

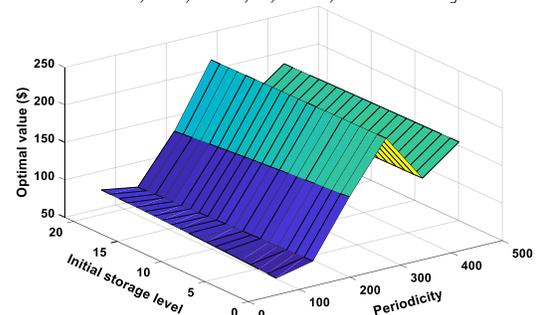

Fig. 6. Optimal values versus periodicity and initial storage levels.

Fig. 6 shows the results. The optimal value is calculated as 1/4-day revenue. For example, if periodicity is selected as one day, we calculate the average revenue of one day for the whole data horizon and divide the average revenue by 4 to obtain 1/4-day revenue. As periodicity changes, the optimal value changes largely, which is mainly because the real-time price is volatile. Fig. 6 shows that setting one day as the planning horizon gives the best revenue. The reason is that our data show that in most cases, the real-time prices in the first half of the day are low, while the prices in the second half are high and volatile; thus, storage tends to charge first and then discharge to earn profits.

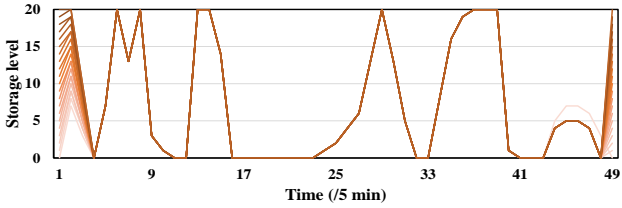

Fig. 7. Storage levels change with different initial storage levels when periodicity is set as 48*5 min.

Moreover, Fig. 6 shows that the influence of initial storage levels on the optimal value is not significant. Fig. 7 further validates this phenomenon. Although initial storage levels vary from 0 to 20, storage levels become the same from the 3rd time point to the 42nd time point. The reason is that optimal control decisions become independent of initial levels after several time steps and will become dependent again with initial levels at the end of the operation horizon, since the final storage level must be equal to the initial storage level. According to (20), optimal decisions are piecewise linear; hence, the mapping from current storage level $X_t$ to the next storage level $X_{t+1}$ is piecewise linear with coefficients not greater than 1. Therefore, the initial storage level will only manifestly affect storage levels of the first few time periods.

In the following discussion, we choose the initial storage level as 0 and periodicity as 4*72*(5 min). Those parameters provide the best revenues under the selected data.

### B. Price maker versus price taker

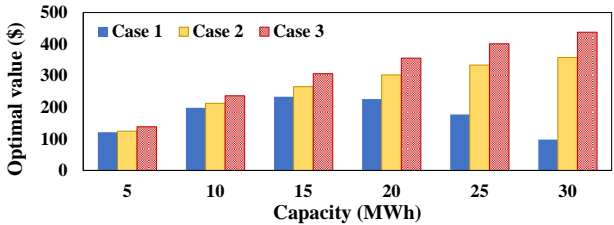

Fig. 8. Optimal values versus capacity under Cases 1, 2 and 3.

Fig. 8 compares optimal values versus capacity in Cases 1, 2 and 3. From Case 1 and Case 2, it can be concluded that the price maker gains more profits than the price taker. This is consistent with our analytical analysis: as the price taker neglects the impacts of its operations on market prices, it will misestimate the actual income during optimization. Moreover, as the storage capacity increases, the price-taker assumption becomes even more impractical. In the price-maker scenarios, i.e., Case 2 and Case 3, optimizing both energy arbitrage and social welfare brings more benefits than merely considering energy arbitrage.

### C. Energy arbitrage gains versus social welfare

Fig. 9 shows the optimal actions under Cases 2 and 3. Different layouts come out in the two cases. In general, storage tends to charge when the market price is predicted to be low and discharge when the price is forecasted to be high. From Fig. 10, we can see that storage acts as a filter that smooths the price pattern and reduces arbitrage opportunities.

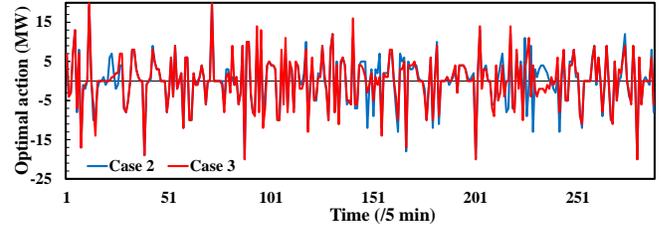

Fig. 9. Optimal actions under Cases 2 and 3.

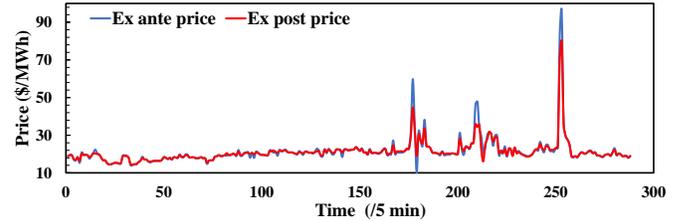

Fig. 10. Price changes under Case 3. Ex ante price and ex post price refer to the market price before and after charging/discharging, respectively.

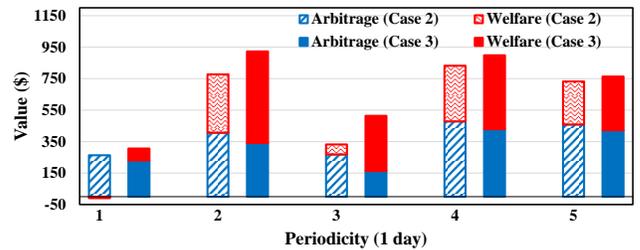

Fig. 11. Optimal values (profits summation of arbitrage and welfare) with time variations under Cases 2 and 3.

In addition, we analyzed arbitrage gains and welfare gains in Cases 2 and 3, respectively, as shown in Fig. 11. Although price-smoothing effects due to storage use reduce the arbitrage value, gains from social welfare enhance the value of energy storage. Note that even if Case 2 does not optimize social welfare, the social welfare still changes because it is a byproduct from arbitrage. Along the period, the total value, i.e., profits summation of arbitrage and welfare, in Case 2 is always smaller than that in Case 3 since the objective in Case 3 is exactly the total value, yet that in Case 2 it is a part of the total value (i.e., the arbitrage part). If the storage belongs to a merchant, profit-maximizing purpose in Case 2 is a good choice. However, if the storage is owned by prosumers/consumers/producers, welfare-maximizing purpose in Case 3 will help gain more profits, which may exceed 40%.

### D. The proposed algorithm versus the SDP algorithm

In this subsection, we investigated the performance of the proposed algorithm and the SDP algorithm in terms of optimality and computational efficiency under the variation of capacity $C$. Set the value of $N_{SOC}$ to $5C$.

The optimality and computational efficiency of the algorithm are illustrated in Fig. 12. The figure shows that the proposed threshold-based algorithm achieves the same optimal value as the SDP algorithm. However, the proposed algorithm is much faster than the SDP algorithm. As the capacity increases, the computation cost of SDP increases exponentially, while that of the threshold-based algorithm grows much slower. The computational burden of the SDP algorithm is mainly due to the high granularity of state discretization. To some extent, the proposed algorithm avoids state discretization, hence accelerating the execution process.

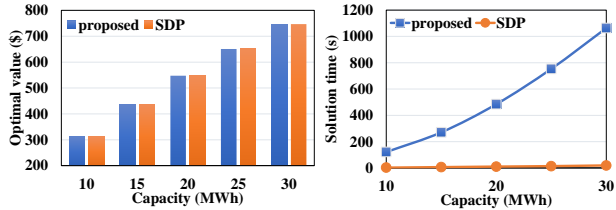

Fig. 12. Comparisons of the proposed algorithm and the SDP algorithm with capacity variations: (a) optimal value (b) solution time.

*E. Sensitivity analysis and optimal capacity*

Fig. 13 shows the change in optimal values with storage capacity and efficiency. We can see that as capacity increases, expected profits grow fast at the beginning but tend to be saturated eventually. Furthermore, storage will earn more profits if it has higher efficiencies.

In addition, we can derive the optimal capacity from Fig. 13. Suppose that the capital cost of the storage is $\rho$. Adding the investment cost to the objective function, we have

$$\max_{C} \ U(C) - \rho C \tag{22}$$

where $U(C)$ is the value-capacity function plotted in Fig. 13 under a given efficiency. Taking the derivative of (22), we have $U'(C^*) = \rho$. Thus, the optimal capacity $C^*$ can be obtained.

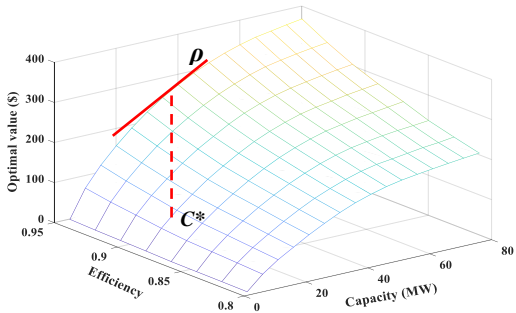

Fig. 13. Optimal values versus storage capacity and efficiency.

## VI. CONCLUSIONS

In this paper, we have proposed a scheme to optimally schedule communal energy storage jointly owned by prosumers in the real-time market. We show the following:

1) The arbitrage behaviors of large-scale energy storage will significantly impact market prices and social welfare. Therefore, for prosumer-oriented communal energy storage, it is important to pursue both energy arbitrage gains and social welfare gains to obtain more benefits.

2) The proposed threshold-based structure is appealing to decision makers due to its ease of implementation and efficient computation. The structure has a simple form to make optimal decisions by comparing the current marginal value with the expected future marginal value.

The proposed method can guide prosumer-based communities to utilize energy storage for profit making. Aside from energy arbitrage and social welfare, participating in auxiliary services accounts for another major source income for storage, which will be characterized in future work. In addition, uncertainty modeling of renewable sources and market prices is another important direction to study.